\documentclass[12pt, draftclsnofoot, onecolumn]{IEEEtran}
\usepackage{amsmath,amsfonts}
\usepackage{algorithmic}
\usepackage{algorithm}
\usepackage{array}
\usepackage[caption=false,font=normalsize,labelfont=sf,textfont=sf]{subfig}
\usepackage{textcomp}
\usepackage{stfloats}
\usepackage{url}
\usepackage{verbatim}
\usepackage{graphicx}
\usepackage{cite}
\usepackage{multirow}
\usepackage[binary-units]{siunitx}
\begin{document}

\title{The Confluence of Blockchain and 6G Network: Scenarios Analysis and Performance Assessment}

\author{Bo Li,
        Shuiguang Deng,~\IEEEmembership{Senior Member,~IEEE}, Xueqiang Yan, and Schahram Dustdar,~\IEEEmembership{Fellow,~IEEE}
          
\IEEEcompsocitemizethanks{\IEEEcompsocthanksitem
            B. Li and S. Deng 
          are with the College of Computer Science and Technology, Zhejiang University, Hangzhou 310027, China. E-mail:  \{bol, dengsg\}@zju.edu.cn. 
      \IEEEcompsocthanksitem X. Yan is with 2012 Lab, Huawei Technologies, Shenzhen 201206, China. E-mail: yanxueqiang1@huawei.com.
          \IEEEcompsocthanksitem S. Dustdar is with Distributed Systems Group, TU Wien, 1040 Vienna, Austria. E-mail: dustdar@dsg.tuwien.ac.at.}

           \thanks{This work was partially supported by the National Science Foundation of China (NSFC) under Grants U20A20173 and 62125206. Schahram Dustdar’s work is also supported by
the Zhejiang University Deqing Institute of Advanced Technology and Industrialization (ZDATI).}
             \thanks{(Corresponding author: Shuiguang Deng)}}

\markboth{ IEEE TRANSACTIONS ON WIRELESS COMMUNICATIONS}%
{Shell \MakeLowercase{\textit{et al.}}: A Sample Article Using IEEEtran.cls for IEEE Journals}


\maketitle

\begin{abstract}

Emerging advanced applications, such as smart cities, healthcare, and virtual reality, demand more challenging requirements on sixth-generation (6G) mobile networks, including the need for improved secrecy, greater integrity, non-repudiation, authentication, and access control. While blockchain, with its intrinsic features, is generally regarded as one of the most disruptive technological enablers for 6G functional standards, there is no comprehensive study of whether, when, and how blockchain will be used in 6G scenarios. Existing research lacks performance assessment methodology for the use of blockchain in 6G scenarios. Therefore, we abstract seven fine-grained 6G possibilities from the application layer and investigate the why, what, and when issues for 6G scenarios in this work. Moreover, we provide a methodology for evaluating the performance and scalability of blockchain-based 6G scenarios. In conclusion, we undertake comprehensive experimental to assess the performance of the Quorum blockchain and 6G scenarios. The experimental results show that a consortium blockchain with the proper settings may satisfy the performance and scalability requirements of a 6G network.

\end{abstract}

\newpage


\section{Introduction}

\IEEEPARstart{W}{ith} the emergence of 5G networks, we have entered a new age of digital society in which various smart applications, including industrial automation, intelligent transportation, and remote healthcare, are thriving \cite{chettri2019comprehensive}. The enormous rise of mobile traffic, which is projected to reach $\SI{607}{\exa\byte}$ by 2025 and 5,016 Exabyte/month by 2030 \cite{hewa2020role}, renders 5G incapable of meeting the new needs of future significant applications. Moreover, the fast growth of data-centric intelligent systems reveals new latency constraints of 5G networks \cite{nguyen20216g}. Thus, several research programs are transitioning towards the next generation of mobile networks, etc., 6G, with the goal of satisfying increasingly severe requirements such as latency, connection, scalability, and reliability by combining diverse networks spanning space, air, and ground \cite{khan2021blockchain,maksymyuk2020blockchain}.

Compared to its 5G predecessor, 6G is anticipated to be an ubiquitous integrated network with faster transmission speed, lower communication latency, improved dependability, and larger coverage. Despite the fact that the emergence of advanced technologies, such as edge intelligence, TeraHertz (THz) communication, wireless optical technology, and large-scale satellite constellation, promotes the implementation of 6G, there are still a number of obstacles to overcome prior to the actual landing. Generally speaking, the issues encountered by 6G may be divided into two categories based on the application needs \cite{khan2021blockchain}. The first category includes scalability, latency, throughput, and synchronization, which are performance requirements resulting from future systems' vast interconnectedness. The second category includes security-related requirements such as confidentiality, integrity, non-repudiation, authentication, and access control. The first group permits widespread communication, whilst the second group ensures the security of entities and data transferred.

Recently, blockchain has attracted a great deal of interest from industrial and academic organizations throughout the world. Blockchain is a distributed ledger system that uses consensus algorithms to store chain-structured data consistently and smart contracts to automate operations. Xu et al. \cite{xu2020blockchain} have shown that by incorporating the trustless and automated capabilities of blockchain, resource management and sharing in 6G networks can be made more performance-effective. Thus, blockchain offers a viable option for addressing the second group of challenges described above. In addition, blockchain is lauded for its intrinsic properties, such as decentralization, traceability, anonymity, immutability, and security \cite{cheng2021blockchain}. It is not difficult to deduce that the second set of obstacles may also be addressed by establishing a communication network using blockchain as its underlying technology. Consequently, blockchain is generally regarded as one of the essential 6G enabling technologies.

Despite the above-mentioned capabilities, scalability is a significant hurdle to the widespread use of blockchain from the standpoint of storage and distribution \cite{jiang2021road}. The maintenance of network consistency necessitates that each blockchain node keep a copy of the whole ledger locally, and the blockchain's trustworthiness is maintained by verifying each transaction and block, at the sacrifice of transaction performance. Moreover, blockchain implementation must contend with the Impossible Trinity, i.e. security, decentralization, and scalability. Any two attributes that are realized must come at the price of the third. Therefore, if the blockchain is included into 6G in an irresponsible manner, not only will it not provide any advantages, but it may also pose certain problems. This concern makes it crucial to verify the requirement and efficacy of the integration architecture by conducting a comprehensive analysis of the performance and possible bottlenecks in the blockchain-enabled 6G network. Several studies are currently investigating blockchain integration for 6G, with the majority focusing on addressing specific issues, such as spectrum sharing, service-level agreement (SLA) management, and mobile user privacy protection \cite{xu2020blockchain, refaey2019blockchain, velliangiri2021blockchain}. Although these works have confirmed the advantages blockchain may bring in, a problem-specific integration architecture cannot provide a general guideline for blockchain deployment as a fundamental component of the 6G network. To the best of our knowledge, there are currently few publications addressing the rationale for the integration of blockchain in 6G in terms of foreseen 6G scenarios. Detailed performance evaluations to forecast possible integration architectural constraints are also lacking. To fill this gap, we present a comprehensive perspective by studying and assessing the role of blockchain in seven plausible 6G scenarios. In addition, we propose a methodology for evaluating the performance and scalability of blockchain-based 6G scenarios . Finally, we implement it in a real - life environment to undertake a thorough assessment of its performance. This paper is intended to serve as an enlightening guideline to spur interest and further investigations for subsequent research on blockchain-empowered 6G systems. The main contributions are summarized as follows:
\begin{itemize}
    \item We extract seven fine-grained scenarios from the foreseeable 6G application layer to analyze whether, when, and how to integrate blockchain into 6G network architecture.
    \item In addition, we propose a methodology for assessing the scalability and performance of blockchains in 6G scenarios.
    \item We conducted performance evaluations on a real network environment consisting of multi-site data centers, on which a consortium blockchain (Quorum) has been implemented. Several configurations, including the number of nodes, compute capacity, and consensus procedure, are used to evaluate the performance of Quorum. Besides, we conducted a more extensive performance experiment based on the Poisson distribution's transaction arrival model. The solid experimental results indicate that blockchain can be integrated into 6G networks with the proper setups.
\end{itemize}
The remainder of this paper is organized by the following order: Section II presents a review of related works. Analyses on detailed 6G scenarios are presented in Section III. The methodology is described in Section IV. In Section V, we illustrate the extensive experimental evaluation. Finally, we conclude the work in Section VI.

\section{RELATED WORK}

Since massive data connectivity is essential for the ever-increasingly intelligent, automated, and ubiquitous digital world \cite{giordani2020toward}, 6G is gradually developing towards a marginal and distributed structure. However, the huge risks of attacks and threats occurring in a distributed system make it a tough challenge to achieve a high degree of security and privacy in 6G networks. Moreover, how to perform efficient and reliable data management in the 6G data systems such as vehicular data sharing, medical data storage, and access control is a critical but troublesome issue.

Blockchain is a decentralized, immutable, and autonomous database that supports the establishment of trust relationships between untrusted subjects in distributed environments. Many superior characteristics of blockchain, including decentralization, traceability, anonymity, and immutability, make it a promising candidate for integration into the security and data management provisions of the 5G/6G system \cite{khan2021blockchain}. 

Blockchain technology offers some key opportunities in 5G networks, such as Infrastructure for Crowdsourcing, Infrastructure Sharing, International Roaming, Network Slicing, Management, and Authentication\cite{9024627}. But, 5G considers the issue of smooth interoperability between different blockchain platforms. These several limitations can be mitigated in 6G by using consensus algorithms, applying novel blockchain architecture and sharing techniques, and increasing the block size of the network\cite{9144301}.

Regarding security issues arising from heterogeneous standard integration and access delegations in 6G environments, Manogaran et al. \cite{manogaran2020blockchain} introduced a blockchain-based integrated security measure for providing secure access control and privacy preservation for resources and users. Although the performance of the proposed solution is verified by several metrics, the latency caused by block validation in the blockchain has not been studied, nor has the evaluation of the data leakage probability. Deb et al. \cite{deb2020magnum} integrated blockchain into fog nodes and centralized servers to establish a secure model-sharing platform in a 6G-based industrial Internet of Things (IIoT). Besides, some works dove deeper into the field of blockchain-enabled resource sharing and spectrum management in 6G and verified that the integration between wireless networks and the blockchain would allow the network to monitor and manage spectrum and resource utilization in a more efficient manner \cite{xu2020blockchain, liu2021blockchain, hu2021blockchain}. These efforts envision blockchain-based resource management, spectrum sharing, and energy trading as drivers for future 6G use cases.

Although these studies have highlighted the integration of blockchain for 6G, most of them deal with specific issues such as data management \cite{manogaran2020blockchain,asheralieva2019distributed}, spectrum sharing \cite{liu2021blockchain,hu2021blockchain}, and privacy protection \cite{velliangiri2021blockchain,nguyen2020privacy,yang2021blockchain}. The exact scope of requirements may vary in different 6G application scenarios due to the diverse nature of involved entities, such as wearable devices, edge servers, and base stations. Therefore, a comprehensive view of blockchain integration in foreseeable 6G scenarios is of great importance. Besides, the inherent scalability-related issues in blockchain, such as throughput and storage bottlenecks, may become potential threats that hinder the efficient operation of 6G systems \cite{xu2020blockchain}. Thus, deep performance evaluation is vital for further exploration of incorporating blockchain in 6G networks.

To explicitly highlight the unique features and technical requirements of 6G, some surveys present representative applications and shed light on fundamental technologies that are expected to empower future 6G networks \cite{nguyen20216g, jiang2021road, chowdhury20206g}. However, they focused on how blockchain can benefit these applications without delving into integration details such as whether to use blockchain, how to define a transaction, and when to generate a transaction.

To the best of our knowledge, no work has been done to investigate in detail how to integrate blockchain into 6G networks from a general scenario perspective, and there are no methods to review the performance and scalability of blockchain-based 6G scenarios. This paper intends to fill this gap and serve as an enlightening guideline to spur deeper investigations for subsequent research on blockchain-empowered 6G networks.

\section{Scenarios analysis of integrating Blockchain in 6G network architecture}

In this section, we extract seven fine-grained scenarios from emerging 6G applications to conduct a detailed analysis on whether, why, and how to integrate blockchain technology into 6G network architecture. The scenario analysis reflects the actual requirements since it reflects the actual needs of 6G applications, based on which we can judge the performance demands of the blockchain and adjust integration schemes.

\subsection{Public Key Management}

In the 6G era, a huge number of devices are connected for data interaction. Public key encryption schemes inherently need to prevent malicious attacks on devices and exchanged data, such as man-in-the-middle attacks and eavesdropping. Key management is the foundation of all security mechanisms. They do everything from data encryption and decryption to authentication, authorization, and access control. Any compromise of cryptographic keys can lead to compromise for the entire security infrastructure, allowing attackers to decrypt sensitive data, authenticate themselves as privileged users or give themselves access to unauthorized information. Therefore, proper management of public keys is an integral part of the 6G network. As a distributed platform, blockchain has been one of the most viable solutions for storing user keys, and the tamper-proof nature of blocks can be leveraged to build a chain of trust for public keys. There are two use cases for public key management in 6G, public key management for users and public key management for network devices.

\begin{itemize}
    \item The public key management of users is for individual users, for example, users need to add or delete public key information in the blockchain when they register or cancel their devices. When using a public key to encrypt personal information, it not only prevents confidential information from being stolen, but also well meets the requirements of GDPR. At the same time, user authentication and access control are of great significance to ensure a secure network cooperation environment.
    \item The public keys of network devices can be mutually authenticated in multiple network devices, not only to prevent pseudo base stations but also to establish a shared network by different operators.
\end{itemize}

\subsection{ID Management}

One of the major challenges facing 6G network operators is bringing all parties together and coordinating their efforts to provide economically viable and seamless connectivity to users. For each new participant, the demand for interfaces with secure authentication and authorization mechanisms will increase, along with the complexity and operational costs of the ID infrastructure required for the associated identity management. While today's centralized ID infrastructures have proven to be technically feasible in limited and trusted spaces, once centralized identity providers must be avoided and due to limited cross-domain interoperability or national data protection legislation and certification, they are unable to provide the required security for country-dependent institutions typically cannot be trusted, for example, geopolitical reasons\cite{garzon2021decentralized}.

A blockchain-based 6G network enables secure mutual authentication across networks with different trust domains. It also allows the network to be independent of trusted third parties while improving the auditability and transparency of IDs. Better management of IDs in multiple trust domains. two use cases for ID management in 6G networks are Pseudo-name management and decentralized ID management.

Pseudonyms, as a data protection method strongly recommended by GDPR, emerge to prevent real information leakage. Using real public keys to create pseudonyms and recording the mapping relationship between pseudonyms and real public keys in the blockchain ensures both the leakage of real public key information and the authenticity of public keys and the auditability of related user behavior.

\subsection{Authorization, Authentication, and Access Control}

In 6G networks, the total number of devices is growing at an increasing rate, which poses new security risks and challenges to the system. Failure to protect network devices from unauthorized access can often lead to serious data breaches, as these devices often contain large amounts of valuable and sensitive data. As for traditional access control technology, centralized management can lead to data leakage, as well as the difficulty of coordinating multiple parties as multiple organizations are involved. Therefore, it is not applicable to 6G networks. Therefore, it is not applicable to 6G networks. Authentication, Authorization, and Access Control(AAA) can be ported to blockchain networks and, in particular, be implemented as a smart contract on a decentralized blockchain with no downtime, no fraud, and no third-party intervention. It also enables secure authentication, authorization, and access control in mutually untrusted administrations.

\subsection{Context information management}

With the objective of providing high quality of service (QoS), 6G system will need to be context-aware i.e., use context information in a real-time mode depends on network, devices, applications, and the environment of users\cite{alam2016towards}. There are several benefits to using blockchain to preserve context information. First, it enables easy access. Because different kinds of context information are kept in the blockchain, there is no need to go through a third-party platform.Secondly, all the modification and deletion records of the context information can be audited, thus enhancing the security of the information. We propose two use cases for context information as follows:

\begin{itemize}
    \item Personal context information is indexed by the user's identity and contains personal information. For example, the cached information such as ID and public key generated by individual users. Putting the personal context information into the blockchain can facilitate the base station to access the cached information quickly and also ensure the auditable record of information usage, thus protecting the privacy of individual users.
    \item Location information is very important context information, which can ensure that operators can better serve their customers. By storing location information in the blockchain, it can facilitate fast access by different operators. However, location information is private information, it needs to be placed in the blockchain by encryption, and access to location information needs to be approved by the owner.
\end{itemize}

\subsection{Data Management and Data Trading}
As digitization accelerates, every element of society is generating large amounts of data all the time, and in turn, benefits from the proliferation of data \cite{liu20216g}. As a result, data management and further data transactions have become one of the key technical building blocks of the 6G architecture. Considering that 6G is envisioned to assume an important role in enabling large-scale IoT devices to seamlessly collaborate to meet highly diverse business needs and to realize the vision of ubiquitous AI. In this paper, we mainly consider subscription data, AI model data, IOT data, and sensing data. all of which contain a large amount of private information and These data contain a large amount of private information and are of high commercial value. Therefore, there is an urgent need for secure systems that support data management and transactions. Blockchain is a distributed database maintained by multiple parties, and it is transparent, traceable, collaboratively maintained, and supports the flow of data and transactions without security issues. With these advantages, blockchain has emerged as one of the potential solutions for data management and transactions.

\begin{figure}[!htbp]
  \centering
  \includegraphics[width=\linewidth]{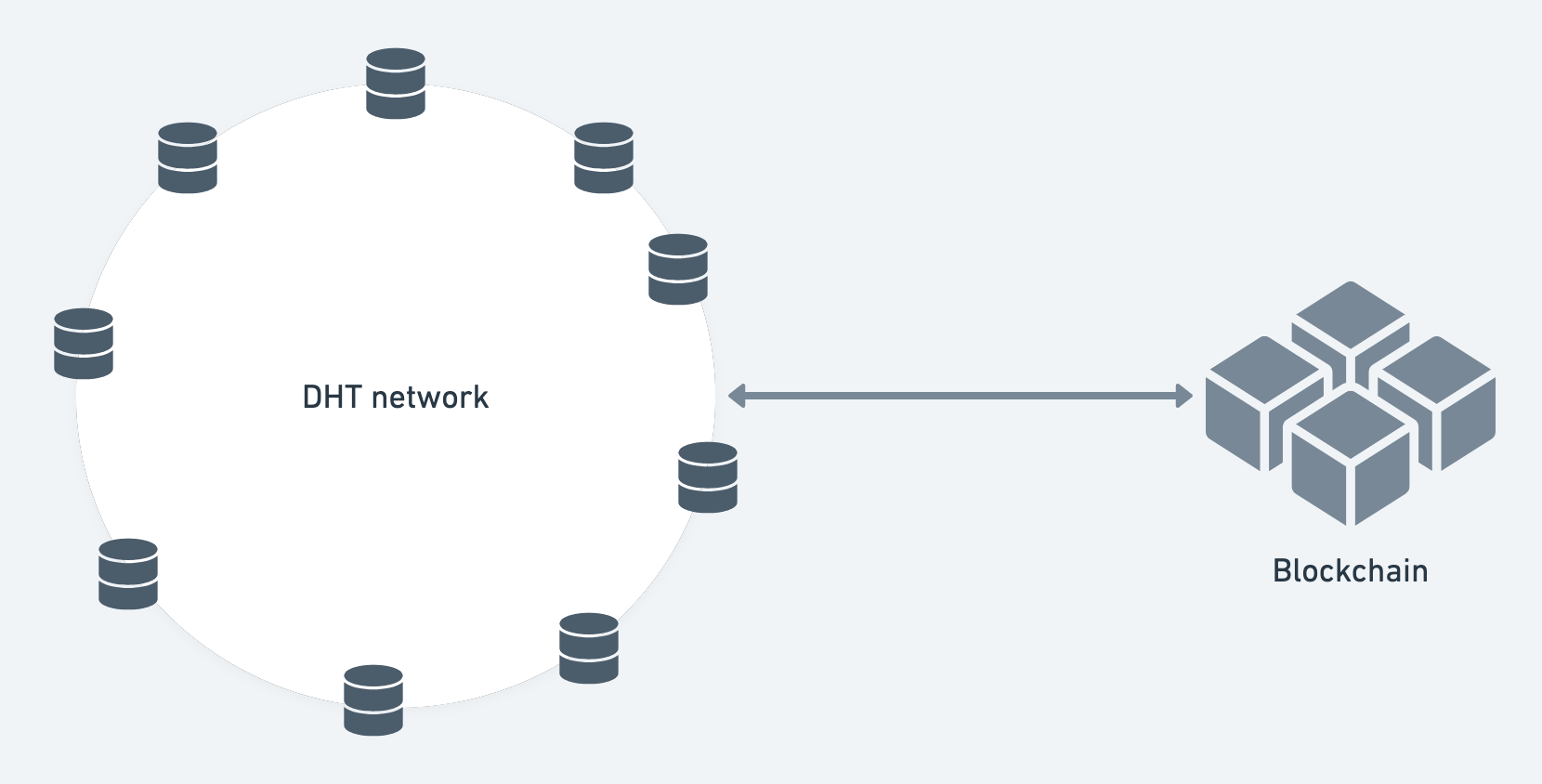}
  \caption{ DLT+DHT architecture}
  \label{fig:DHT_DLT}
\end{figure}

We use both on-chain and off-chain architectures to manage data, as shown in \ref{fig:DHT_DLT}. The hash address of the original data is stored on the blockchain, while the original data is stored off-chain after encryption. For data deletion and update, the data activity should be recorded on the blockchain. The on-chain/off-chain architecture ensures the "right to forget" as required by GDPR and also ensures the expansion of the blockchain ledger due to excessive data size.

\subsection{Resource Sharing}
To achieve the goals of 6G, wireless resources such as spectrum, compute, storage, and infrastructure plays a critical role, and the cost of sharing these resources will be significant. Traditional studies rely on a centralized third party to validate each shared transaction, which is vulnerable to many security threats, including single points of failure, denial of service attacks, etc\cite{8977442}. Moreover, they focus only on resource management and ignore privacy and security issues that are critical to resource sharing.

Resource sharing is a typical use case where blockchain can be used to efficiently exchange assets between multiple stakeholders without the need for a centralized third party to provide trust. In a blockchain, all resource sharing and transactions are transparent and secure. Not only that, in resource sharing, all shared executions can be consistent through smart contracts without human intervention. 

In 6G networks, there are several resources that can be shared, such as spectrum, computing resources, and networks.

\subsection{Trading and Settlement}

Traditional asset transactions require the involvement of intermediaries, such as brokers or paying agents, to facilitate the clearing and settlement of transactions, making the settlement process very time-consuming and costly. Blockchain can be used to exchange assets between multiple stakeholders in a de-trusted environment. as it provides a decentralized infrastructure and enables more flexible settlement cycles to speed up settlements. Through smart contracts, we can automate transactions and settlements without human intervention to ensure the security of assets and ease of transactions.

At the same time, due to the in-mutability of the blockchain, all transaction and settlement information can be accessed through the blockchain, which can also facilitate future inquiries and audits. In 6G, we propose two use cases that require the use of blockchain.

\begin{itemize}
    \item In the wireless telecommunications environment, there is interconnection settlement, roaming settlement, and billing between different operators. The existing settlement methods take a long time and the results are not clear and ambiguous. With the introduction of blockchain, different stakeholders can transact and settle faster, more transparently, more accurately, and more securely.
    \item  Call Detail Records (CDRs) are used to charge customers for using transportation network services at the end of the billing period. To make billing information auditable, CDRs can be periodically recorded on the blockchain or stored in an on-chain/off-chain architecture (Figure \ref{fig:DHT_DLT}). When settlement and transactions are encountered, the CDRs are queried via smart contracts to automatically perform billing and roaming tasks.
\end{itemize}

\begin{table}[]
\caption{6G Scenario Analysis Based on Blockchain:Scenario 1-4 (W for Write Transaction)(R for Read Transaction)}  
\label{table1}
\resizebox*{\textwidth}{!}{
\begin{tabular}{|p{2.5cm}|p{3cm}|p{4.5cm}|p{3.5cm}|p{4cm}|}
\hline
& Use cases & Why on-chain? Benefits? & 
\begin{tabular}[c]{@{}l@{}}What is recorded on \\ chain? i.e. transaction \\ definition
\end{tabular}& 
\begin{tabular}[c]{@{}l@{}}When is the \\ transaction \\ generated/query?
\end{tabular}                                                                             \\ \hline
\multirow{8}{*}{\begin{tabular}[c]{@{}l@{}}1 Public key \\ management\end{tabular}}                                   & \begin{tabular}[c]{@{}l@{}} Subscribers’ \\ public key \\ management\end{tabular}         & \begin{tabular}[c]{@{}l@{}}$\bullet$ Decentralization, i.e. avoid \\ centralized PKI;\\ $\bullet$ Tamper-proof ensures the \\ authenticity of public key;\\ $\bullet$ Provide public key to 3rd \\ party to authenticate the \\ user.\end{tabular} & \begin{tabular}[c]{@{}l@{}}$\bullet$ Transaction content: \\ \{Hash(ID) : Public key\}\\ $\bullet$ Transaction is digitally \\ signed by operator’s \\ private key\end{tabular}                                                                               & \begin{tabular}[c]{@{}l@{}}$\bullet$ When an end user \\ subscribes to the \\ network provider.(W)\\  $\bullet$ User query, operator \\ query.                      (R)\end{tabular}                                                                                                                                                        \\ \cline{2-5} 
&\begin{tabular}[c]{@{}l@{}}Network \\ equipment’s\\ public key\\ management\end{tabular} & \begin{tabular}[c]{@{}l@{}}$\bullet$ Decentralization, i.e. avoid \\ centralized PKI;\\ $\bullet$ Tamper-proof ensures the \\ authenticity of public key;\\ $\bullet$ Provide public key to 3rd \\ party to authenticate the \\ user.\end{tabular} & \begin{tabular}[c]{@{}l@{}}$\bullet$ Transaction content:\\ \{Hash(NEID) : Public \\ key\} \\ $\bullet$ Transaction is digitally \\ signed by operator’s \\ private key\end{tabular}                                                                          & \begin{tabular}[c]{@{}l@{}}$\bullet$ When network \\ equipment is onboard.(W)\\ $\bullet$ operator query.            
(R)\end{tabular}  
\\ \hline
\multirow{5}{*}{2 ID  management}
& \begin{tabular}[c]{@{}l@{}}Pseudo-name \\ management\end{tabular}                       & \begin{tabular}[c]{@{}l@{}}$\bullet$ User’s public key is \\ endorsed by the central \\ authority - authenticity.\\ $\bullet$ Auditable users’ behavior\end{tabular}            
& \begin{tabular}[c]{@{}l@{}}$\bullet$ Transaction content: \\ \{pseudo-name: public \\ key\}\\ $\bullet$ Transaction is digitally \\ signed by central \\ authority who creates \\ the pseudo-name\end{tabular}                 
& \begin{tabular}[c]{@{}l@{}}$\bullet$ When the pseudo-name \\ is created by central \\ authority. (W) \\ $\bullet$ When the pseudo-name \\ is queried.(R)\end{tabular}         
\\ \cline{2-5} 
&\begin{tabular}[c]{@{}l@{}} Decentralized ID \\(DID)\end{tabular}                         & \begin{tabular}[c]{@{}l@{}}$\bullet$ publicly accessible, \\ Transparent.\\ $\bullet$ Trusted attestation.\end{tabular}                                                         & \begin{tabular}[c]{@{}l@{}}$\bullet$ identifiers and use \\ schemas\end{tabular}      
& \begin{tabular}[c]{@{}l@{}}$\bullet$ When the DID is created \\ by central authority. (W) \\ $\bullet$ when the identifier is \\ verified.                      
(R)\end{tabular}                                                                        
\\ \hline

\multirow{8}{*}{\begin{tabular}[c]{@{}l@{}}3 Authentication, \\ Authorization, \\ and Access \\ control\end{tabular}} & Authentication& \multirow{15}{*}{\begin{tabular}[c]{@{}l@{}}$\bullet$ Traceable \& auditable records \\of user’s subscription data access \\activities as required by GDPR \\ or PIPL\end{tabular}}                            &\begin{tabular}[c]{@{}l@{}}$\bullet$ The data activity of \\ inquiring the \\ subscriber’s data \\ (subscriptions, profiles)\end{tabular}                              &\begin{tabular}[c]{@{}l@{}}$\bullet$ When a service request \\ is initiated.        (R \& W)\end{tabular}                                                                          \\ \cline{2-2} \cline{4-5} 
& Authorization& & \begin{tabular}[c]{@{}l@{}}$\bullet$ The data activity of \\ inquiring the\\ subscription profile.\end{tabular}                                                                                                                                       & \begin{tabular}[c]{@{}l@{}}$\bullet$ When a specific service \\ request.              (R \& W)\end{tabular}                                                                                                                                                                                                                         \\ \cline{2-2} \cline{4-5} 
& Access Control& & 
\begin{tabular}[c]{@{}l@{}}$\bullet$ The data activity of \\ inquiring the user data \\ which is not include in \\ subscription profile.\end{tabular}& \begin{tabular}[c]{@{}l@{}}$\bullet$ When a 3rd party or a \\ network function \\ access to the user’s \\ data.                  
(R \& W)\end{tabular}                                                                   \\ \cline{1-2} \cline{4-5} 
\multirow{4}{*}{4 Context}                                                                      & \begin{tabular}[c]{@{}l@{}}personal context\\ information context\end{tabular}                              &  & \begin{tabular}[c]{@{}l@{}}$\bullet$ The data activity of \\ inquiring the \\ subscription profile \\ and updating/deleting\\ the context.\end{tabular}                 & \begin{tabular}[c]{@{}l@{}}$\bullet$ When a network \\ function access to \\ user’s subscription \\ data, or update/delete the \\ context.             (R \& W)\end{tabular}  \\ \cline{2-2} \cline{4-5} 
& \begin{tabular}[c]{@{}l@{}}Location \\ information\end{tabular} &                                            & \begin{tabular}[c]{@{}l@{}}$\bullet$ Location information \\ access log.\end{tabular}                                                                                                                                                                 & \begin{tabular}[c]{@{}l@{}}$\bullet$ When a 3rd party or a \\ network function access \\ to the user’s data(R \& W)\end{tabular}                                                                                                                                       \\\hline

\end{tabular}
}
\end{table}

At the end of this section, we analyze the seven 6G scenarios in more detail. In Table \ref{table1}, Table II, we analyze the why, what, and when questions for each scenario use case which gives a clearer and more intuitive presentation of each scenario.

\begin{table}[]
\label{table2} 
\caption{6G Scenario Analysis Based on Blockchain:Scenario 5-7}  
\resizebox*{\textwidth}{!}{
\begin{tabular}{|p{2.875cm}|p{3.45cm}|p{5.175cm}|p{4.025cm}|p{4.6cm}|}
\hline
\multirow{18}{*}{\begin{tabular}[c]{@{}l@{}}5 Data \\ management \& \\ data trading\end{tabular}} & \begin{tabular}[c]{@{}l@{}}Subscription data\end{tabular}& \begin{tabular}[c]{@{}l@{}}$\bullet$ GDPR/PIPL, including \\ ‘forgettable/erasable\end{tabular}& \begin{tabular}[c]{@{}l@{}}$\bullet$ Hash/address of the \\ off-chain stored \\ subscription profile\\ $\bullet$ For removing data: The \\ action of removing the \\ off-chain data\\ $\bullet$ For update: The action \\ of updating the offchain\\ data\end{tabular} & \begin{tabular}[c]{@{}l@{}}$\bullet$ User subscribes to the \\ service provided by \\ network provider.  (R \& W)\\ $\bullet$ User change his/her \\ subscription.          (R \& W)\\ $\bullet$ User de-register his/her \\ service from the.    (R \& W)\\ network provider.\\ $\bullet$ User update his/her \\ subscription.            (W)\end{tabular} \\ \cline{2-5}& AI Model data& \begin{tabular}[c]{@{}l@{}}$\bullet$ Tamper-proof\\ $\bullet$ Defend poison attack\\ $\bullet$ Eliminate SPOF (e.g. the \\ aggregator), mitigate the \\ DDoS attacks (targets at \\ centralized aggregator)\end{tabular}                           & \begin{tabular}[c]{@{}l@{}}$\bullet$ The hash of the model \\ data (on-chain/offchain)\\ $\bullet$ Encrypted model data \\ (on-chain store AI \\ model data)\end{tabular}                                                                                    & \begin{tabular}[c]{@{}l@{}}$\bullet$ When the model \\ training is completed.(W) \\ $\bullet$ When the gradient \\ update is completed. (W) \\ $\bullet$ Retrieve the \\ model/gradient.        (R)\end{tabular}\\ \cline{2-5} 
& IoT data & \begin{tabular}[c]{@{}l@{}}$\bullet$ Privacy preserving, \\ traceable, auditable\\ $\bullet$ Avoid the problem of ‘BC \\ bloat’\end{tabular}& $\bullet$ Hash of raw data& \begin{tabular}[c]{@{}l@{}}$\bullet$ Periodically store the \\ streaming/time series \\ data.(W)\\ $\bullet$ Audit and \\ trading/sharing        (R)\end{tabular}\\ \cline{2-5} 
& Sensing data& \begin{tabular}[c]{@{}l@{}}$\bullet$ Auditability\\ $\bullet$ Data sharing (automatic \\ trading)\end{tabular}& \begin{tabular}[c]{@{}l@{}}$\bullet$ Hash of the sensing \\ data\end{tabular}& \begin{tabular}[c]{@{}l@{}}$\bullet$ Periodically store the \\ streaming/time series \\ data. (W)\end{tabular}\\ \cline{2-5} 
& \begin{tabular}[c]{@{}l@{}}Data \\ trading/sharing\end{tabular}                          & \begin{tabular}[c]{@{}l@{}}$\bullet$ Automatic trading (SC)\\ $\bullet$ Trusted data source\\ $\bullet$ Data cooperation\end{tabular}& \begin{tabular}[c]{@{}l@{}}$\bullet$ Data package \\ exchanged between \\ data owner and data \\ requester.\end{tabular}& \begin{tabular}[c]{@{}l@{}}$\bullet$ Data is shared or \\ exchanged when the \\ trading occurs. (R\&W) \\ $\bullet$ Audit (R)\end{tabular}\\ \hline
\multirow{12}{*}{\begin{tabular}[c]{@{}l@{}}6 Resource sharing\end{tabular}}& Spectrum& \multirow{6}{*}{\begin{tabular}[c]{@{}l@{}}$\bullet$ Automatic settlement (via SC)\\ $\bullet$ Automatic auction (via SC)\end{tabular}}& \begin{tabular}[c]{@{}l@{}}$\bullet$ Spectrum resource \\ status\\ $\bullet$ Geographic \\ information is included\end{tabular}& \begin{tabular}[c]{@{}l@{}}$\bullet$ When available \\ spectrum is published. (W) \\ $\bullet$ Trade deal(W) \\ $\bullet$ Revoke                       (W) \\ $\bullet$ Audit                          (R)\end{tabular}\\ \cline{2-2} \cline{4-5} 
& \begin{tabular}[c]{@{}l@{}}Computing \\ resource\end{tabular}                            && \begin{tabular}[c]{@{}l@{}}$\bullet$ Computing resource \\ status\\ $\bullet$ Geographic \\ information is included\\ $\bullet$ Per device/NF/MEC/DC\end{tabular}& \begin{tabular}[c]{@{}l@{}}$\bullet$ When available \\ computational resource \\ is published. (W) \\ $\bullet$ Trade deal (W) \\ $\bullet$ Revoke       (W) \\ $\bullet$ Audit          (R)\end{tabular}\\ \cline{2-5} 
& \begin{tabular}[c]{@{}l@{}}Network sharing \\(RAN,  CN…)\end{tabular}                 & \begin{tabular}[c]{@{}l@{}}Precise, near-real-time (record), \\trusted automatic settlement (via \\ SC)\\ $\bullet$ Auditable \\ $\bullet$ Tamper-proof\end{tabular}& \begin{tabular}[c]{@{}l@{}} Hash of the following \\ information needs to \\ be recorded on-chain \\ $\bullet$ User’s network usage\\ $\bullet$ NE’s resource provision\\ status\end{tabular}                                         & \begin{tabular}[c]{@{}l@{}}$\bullet$ When settlement \\ occurs (W)\\ $\bullet$ batch Log information \\ (W) \\ $\bullet$ Audit (R)\end{tabular}
\\ \hline
\multirow{9}{*}{\begin{tabular}[c]{@{}l@{}}7 Trading \& \\ Settlement\end{tabular}}& \begin{tabular}[c]{@{}l@{}}Interconnection \\ settlement\end{tabular}                    & \begin{tabular}[c]{@{}l@{}}$\bullet$ Auditable usage\\ $\bullet$ Automatic settlement(via \\ SC)\end{tabular}& \begin{tabular}[c]{@{}l@{}}$\bullet$ Interconnection traffic \\ volume/usage (per \\ hour)\\ $\bullet$ Settlement (per \\ month)\end{tabular}& \begin{tabular}[c]{@{}l@{}}$\bullet$ Periodic (W) \\ $\bullet$ Per-hour/Perday/month (W) \\ $\bullet$ Audit (R)\end{tabular}\\ \cline{2-5} 
& \begin{tabular}[c]{@{}l@{}}Roaming \\ settlement\end{tabular}                            & \begin{tabular}[c]{@{}l@{}}$\bullet$ Auditable usage\\ $\bullet$ Automatic settlement(via \\ SC)\end{tabular}& \begin{tabular}[c]{@{}l@{}}$\bullet$ CDR (periodically \\ record on the chain in \\ a batch)\\ $\bullet$ Settlement (per-user)\\ $\bullet$ Per-day/month\end{tabular}& \begin{tabular}[c]{@{}l@{}}$\bullet$ Periodic settlement (W) \\ $\bullet$ Per-hour/Perday/month (W) \\ $\bullet$ Audit(R)\end{tabular}\\ \cline{2-5} 
& Billing& $\bullet$ Auditable usage& \begin{tabular}[c]{@{}l@{}}$\bullet$ CDR (periodically \\ record on the chain in \\ a batch)\end{tabular}& \begin{tabular}[c]{@{}l@{}}$\bullet$ Per-hour/Perday/month (W) \\ $\bullet$ Periodic settlement(R) \\ $\bullet$ Audit(R)\end{tabular}\\ \hline                         
\end{tabular}
}
\end{table}

\section{Methodology of performance evaluation}
In this section, we introduce a methodology which can evaluate the performance and scalability of blockchain-based 6G scenarios. We divide transactions in the blockchain into "read" and "write", and analyze when "read" and "write" are required in seven scenarios. Based on above, we propose an abstract model for the sending rate as poisson distribution. Finally, we calculate the transaction arrival rates of scenarios.

\begin{figure*}[!htbp]
  \centering
  \includegraphics[width=\linewidth]{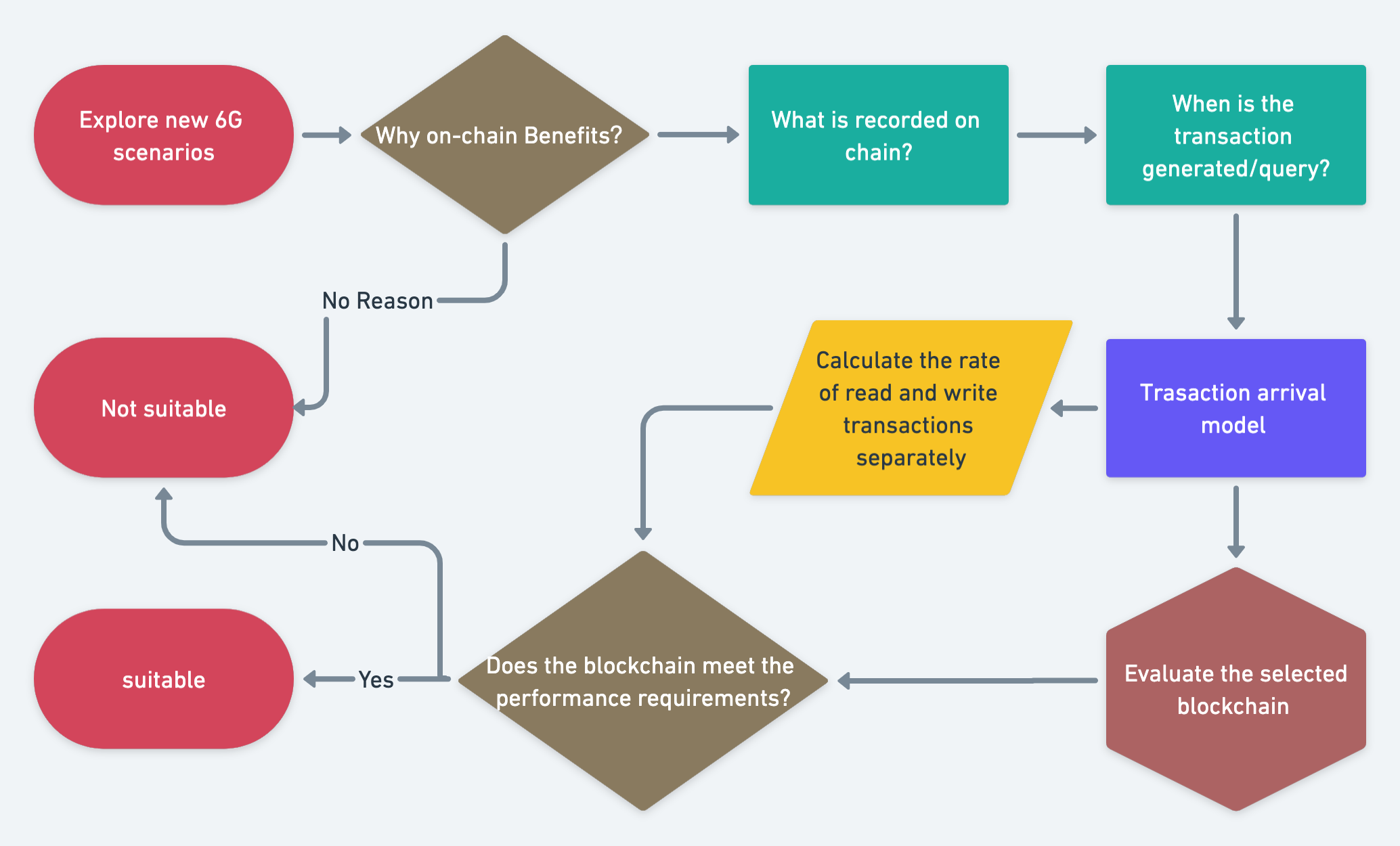}
  \caption{ methodology flow}
  \label{fig:methodology}
\end{figure*}

\subsection{Methodology}

From related work, it can be concluded that there is no common evaluation method for blockchain-based 6G scenarios. In this section, we propose a methodology that can evaluate whether the blockchain performance meets the 6G scenario.

To simplify the evaluation process, we provide the flowchart in figure \ref{fig:methodology}. When you find a new 6G scenario, first you need to understand why the scenario needs to use blockchain and the benefits of using blockchain. If there is no benefit to using blockchain in this scenario, or if the performance and usefulness of the entire scenario are not greatly improved by using blockchain, then the scenario cannot be combined with blockchain. After determining why you want to use blockchain, you should know how to use blockchain in this scenario and what should be recorded on the blockchain. After that, you need to know when to use blockchain. Here, you need to distinguish between "read" and "write" transactions, and you need to know when to "read" and when to "write ".

Next, after determining when to use the blockchain, you need to determine the read and write transaction arrival rate model. Examples include Poisson Distribution Model, Pareto Distribution Process, Weibull Distribution Process, etc. After determining the transaction arrival rate, you need to input the model into the blockchain for performance evaluation and get the "read" and "write" performance of the blockchain.

In Table \ref{table1}-II, we analyze the why, what, and when of the seven scenarios using blockchain. Usually, most 6G scenarios are upgrades of 5G scenarios, so when we consider 6G scenarios, we can extend them by considering some of the 5G scenarios. We were inspired by 5G to calculate the "write" and "read" speeds for such scenarios in 6G.

Finally, you need to compare the calculated "read" and "write" transaction speeds with the maximum "read" and "write" transaction speed of the blockchain. "transactions. If the "read" and "write" performance of the blockchain system meets the scenario, then the scenario can be considered for the blockchain.

\subsection{Transaction arrival model}

With our methodology, in Table \ref{table1}-II, we first analyze what and when to use blockchain in seven scenarios. Different scenarios and different times require different transaction types, so we divide blockchain transactions into "reads" and "writes".

\begin{itemize}

    \item For the sake of brevity, the transaction query operation is referred to as 'read' while the operation of originating and recording.
    
    \item For write operations, we need to change the state of the blockchain and wait for multiple nodes to reach consensus. so write transactions can only be done sequentially.
    
\end{itemize}

Next we need to pick a model for the transaction arrival rate, before that we need to clarify the following concepts.

\begin{itemize}

    \item $\eta$ - The number of concurrent events (CCE) refers to the total number of events that simultaneously occur. Different scenario or use case has different CCE. CCE can be calculated based on some assumptions or can be observed through traffic monitoring on the real network.
    
    \item $\alpha$ - The number of "read" transactions - the number of blockchain transaction query operations performed to complete an event in a given scenario.
    
    \item $\beta$ - The number of "write" transactions - the number of operations to record transactions on the blockchain for completing an event in a given scenario.
    
\end{itemize}

In our case, traffic refers to blockchain transactions proposed or generated by the programs of the different scenarios we analyzed in Section III. For all scenarios and use cases, the Poisson model is a good choice. In this model, the interarrival times have the following characteristics. 

\begin{enumerate}
    \item They are independent. 
    \item They are exponentially distributed, i. e., probability density function.
\end{enumerate}

We can assume that the seven scenarios listed satisfy the above two characteristics\cite{jain1986packet}. Therefore the inter-arrival times are exponentially distributed with a rate parameter $\lambda$:
    
\begin{equation}
    \label{eq : poisson}
    p\left \{ A_n\le \lambda \right \} = 1 - exp(-\lambda t)
\end{equation}

The rate parameter $\lambda$ is determined by the number of the ‘read’ or ‘write’ transactions performed on the blockchain in a specific scenario or use case and the number of CCE.

\begin{equation}
    \label{eq : write}
    \lambda_{write\;transaction} = \eta \times \beta = \lambda_\beta
\end{equation}

\begin{equation}
    \label{eq : read}
    \lambda_{read\;transaction} = \eta \times \alpha = \lambda_\alpha
\end{equation}

\subsection{Case study of transaction arrival analysis}
In this section, we focus on the transaction arrival rate of the 6G scenario.  Assuming a total number of 30 million subscriptions, we can obtain $\lambda_\beta$ and $\lambda_\alpha$ by using the equations \ref{eq : read} and \ref{eq : write}. Since the $\beta$ data comes from existing 5G network operators, here we focus on how many read and write transactions are available for each scenario CCE. Finally, we have selected two typical scenarios to calculate their transaction arrival rates. Due to space reasons, the specific calculations for the remaining scenarios are not given.

\subsubsection{The transaction arrival rate of Public Key Management}
The read operation in public key management is used in the AAA scenario, and the write transaction is mainly described here. The write transaction is mainly when the user opens an account or when a new device goes online (such as a new base station). There is one write transaction per public key managed CCE. (The figure of 0.0015 is from the operator)

\begin{equation}
    \lambda_\beta = \eta \times \beta = 0.0115\times 1 = 0.0115
\end{equation}

\subsubsection{The transaction arrival rate of AAA}

Authentication is mainly about verifying the authenticity of the identity. Usually, a query on the chain is sufficient for authentication, so this scenario is a "read" transaction only. Each authentication requires the user to submit a signature for verification, so only one "read" transaction is required. Authorization is the granting of certain information or rights to another person. It is often necessary to update their authorization information in the blockchain. Therefore, in each Authorization, one "write" transaction is required to change the authorization information, and two read transactions are required to verify the information of the authorized person.

Complete access control requires authorization and authentication. Complete access control requires authorization and authentication. The number of authorizations and authentications required varies from one access control to another, so we refer to the FairAccess framework for evaluation\cite{ouaddah2016fairaccess}.In FairAccess, the complete process requires three authentications and one authorization. (The figure of 8333 is from the operator)

\begin{equation}
    \lambda_\beta = \eta \times \beta = 8333\times 1 = 8333
\end{equation}

\begin{equation}
    \lambda_\alpha = \eta \times \alpha = 8333\times (1 \times 3 + 2 \times 1)  = 41665
\end{equation}

\section{Experimental Evaluation}

\begin{figure}[!htbp]
  \centering
  \includegraphics[width=\linewidth]{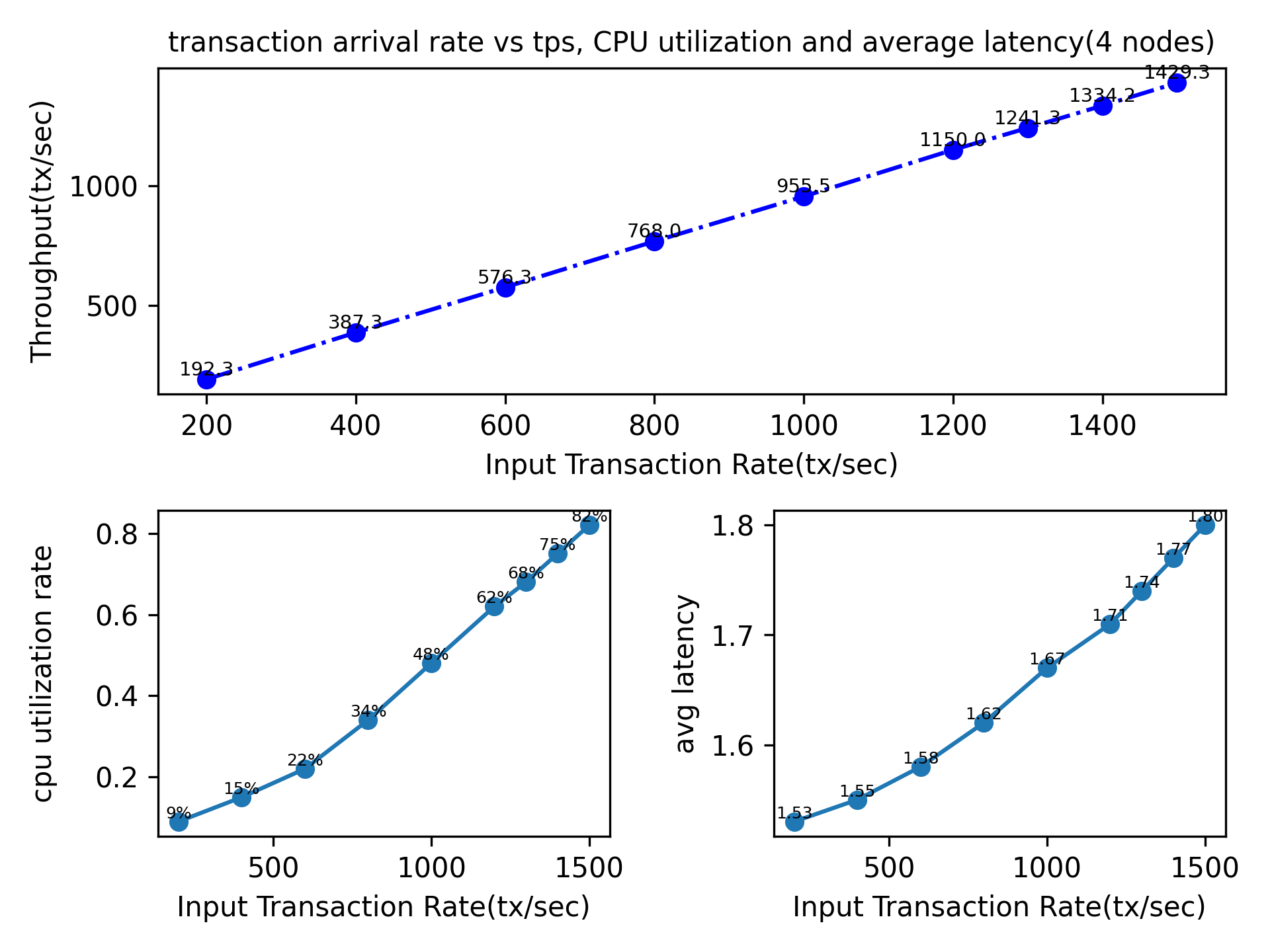}
  \caption{transaction arrival rate vs tps, CPU utilization and average latency(4 nodes write)}
  \label{fig:03}
\end{figure}

\subsubsection{Consensus Quorum}
\begin{figure}[!htbp]
  \centering
  \includegraphics[width=\linewidth]{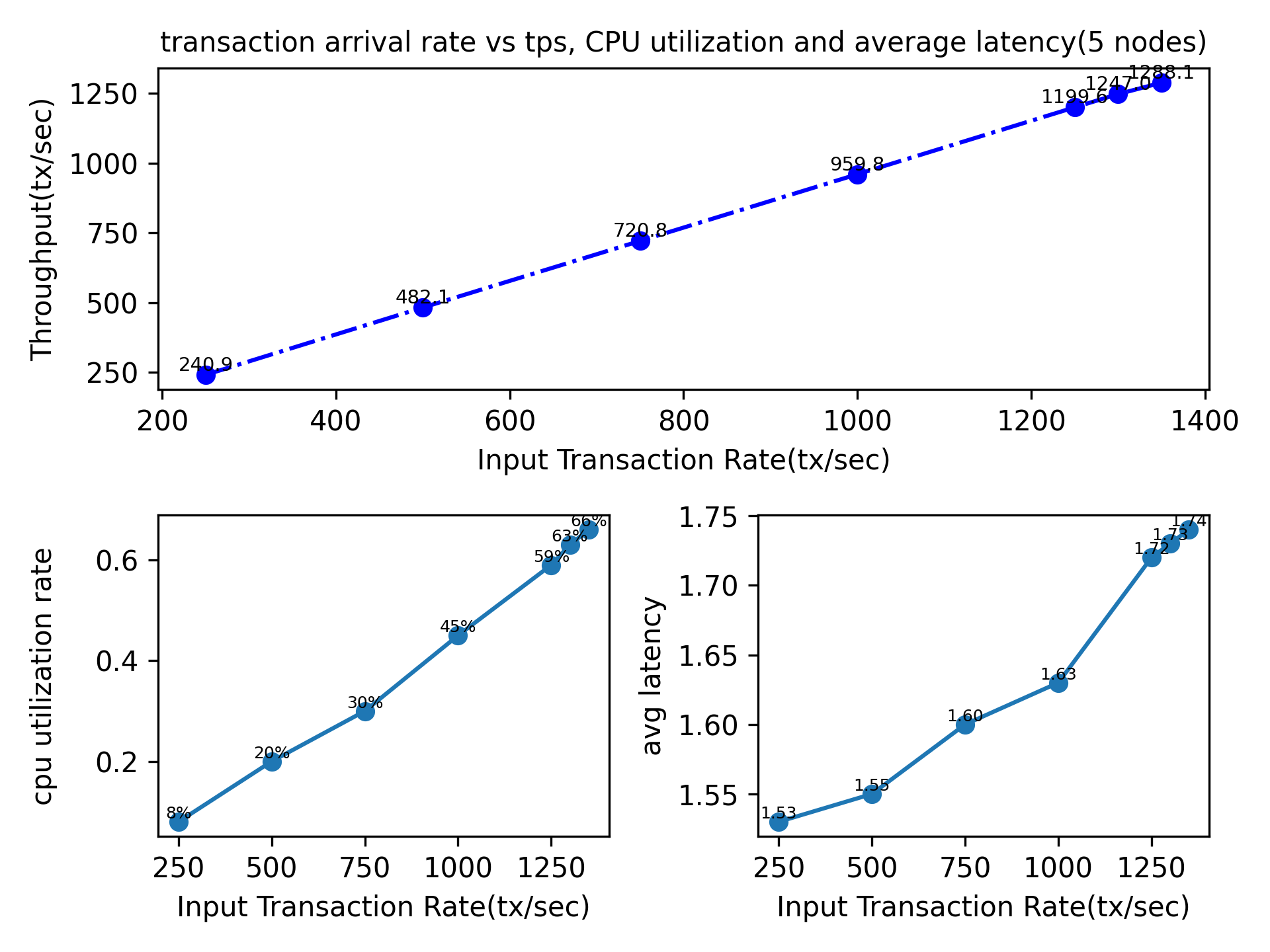}
  \caption{transaction arrival rate vs tps, CPU utilization and average latency(5 nodes write)}
  \label{fig:04}
\end{figure}

Quorum provides unified control for infrastructure management and blockchain network governance. Quorum is compatible with Ether-related components, so we can get better results in our experiments. As shown in Figures \ref{fig:03}-\ref{fig:08}, we also performed a more comprehensive performance and storage evaluation of Quorum and obtained more desirable results.

In Quorum, we chose the BFT-like consensus algorithm IBFT. We measured the size of empty blocks in Quorum and also compared the block time and storage relationship. In Figures \ref{fig:03}-\ref{fig:06}, we measured the performance of write transactions in 4 nodes and compared the number of nodes with the performance of write transactions.

\subsubsection{Cloud Setup}

In each experiment, N consensus nodes are deployed in Cloud virtual machine. Each virtual machine is equipped with a 4-core 8-thread 2.6GHz CPU, 16GB RAM, and 100GB storage. The round-trip latency between any two virtual machines is about 30ms. These cloud servers are deployed in Shanghai, Beijing, Guangzhou, Guiyang, and Ulanchabu respectively.

\subsection{Performance Metrics}
In the literature, we are mainly concerned with the latency, performance, and resource costs of the system. At the same time, we will separate "read" and "write" and measure their performance separately. We divided into two groups of experiments, namely "read" and "write" performance evaluations. 

\begin{itemize}
    \item Latency: Mainly refers to the average time of each transaction from when it is sent to when it is completed.
    \item Throughput: This indicates two metrics, namely a) read transaction throughput and b)write transaction throughput.
    \item resource Costs: Our main consideration is memory consumption and computational consumption.
\end{itemize}

\begin{figure}[!htbp]
  \centering
  \includegraphics[width=\linewidth]{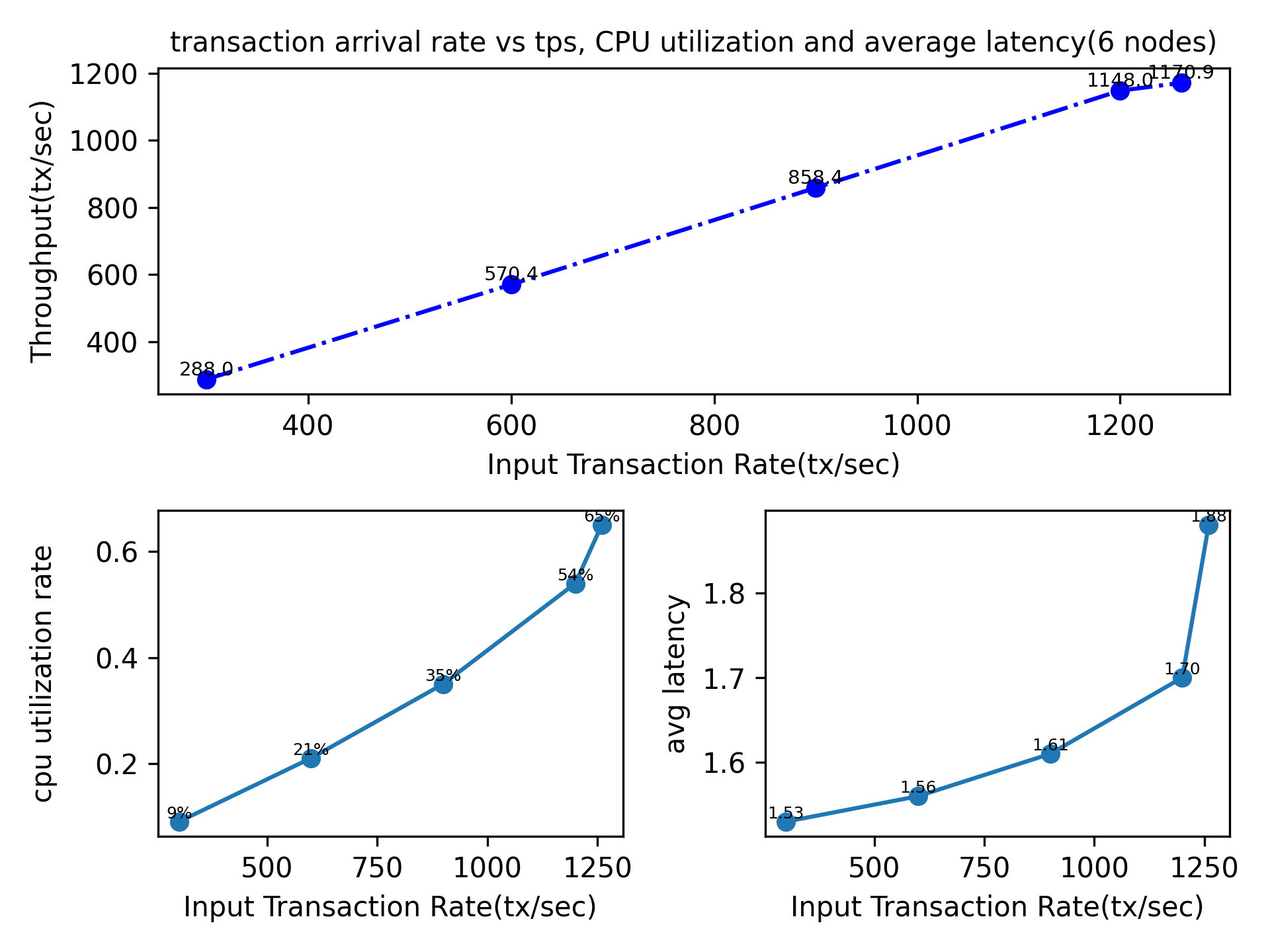}
  \caption{transaction arrival rate vs tps, CPU utilization and average latency(6 nodes write)}
  \label{fig:05}
\end{figure}

\subsection{Performance Assessment}
We use caliper to generate read and write transactions, and simulate transaction generation of read and write transactions for the system respectively.

\subsubsection{Quorum Basic Performance Assessment} 

\begin{figure}[!htbp]
  \centering
  \includegraphics[width=\linewidth]{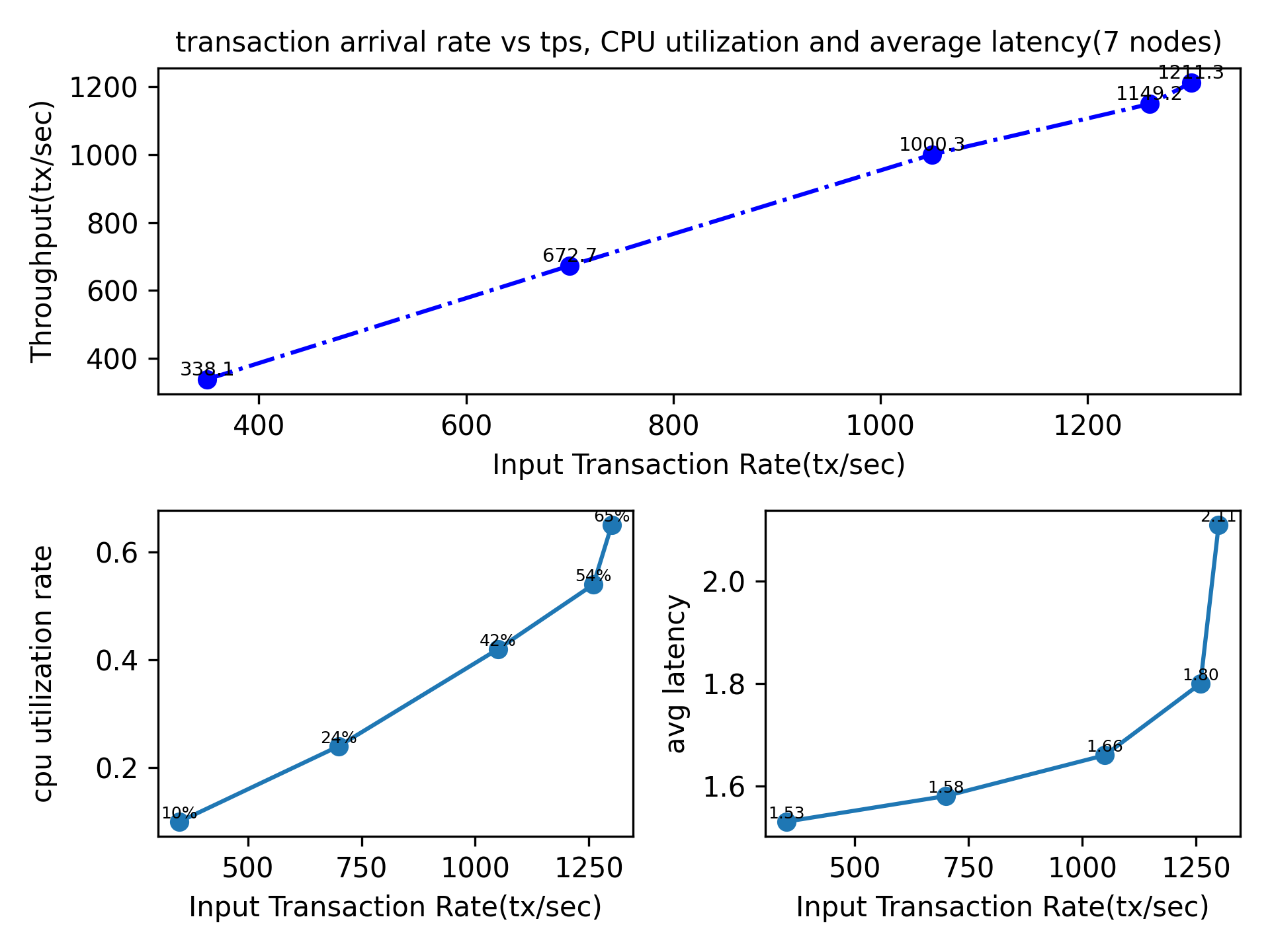}
  \caption{transaction arrival rate vs tps, CPU utilization and average latency(7 nodes write)}
  \label{fig:06}
\end{figure}

First, as shown in Figures \ref{fig:03}-\ref{fig:06}, we evaluated the write performance of the blockchain with 4, 5, 6, and 7 nodes. From Figures 3-6 we find that the TPS of the system grows linearly until the transaction arrival rate reaches the maximum. But as the transaction rate reaches its peak, the system starts to lose steady-state (we define steady-state as the transaction arrival rate of the system = system throughput rate) and the TPS starts to gradually decline. At the same time, CPU usage and latency begin to grow as the arrival rate increases. we find that the maximum TPS becomes smaller and smaller as the number of nodes grows. This is the result caused by the BFT algorithm.

\begin{figure}[!htbp]
  \centering
  \includegraphics[width=\linewidth]{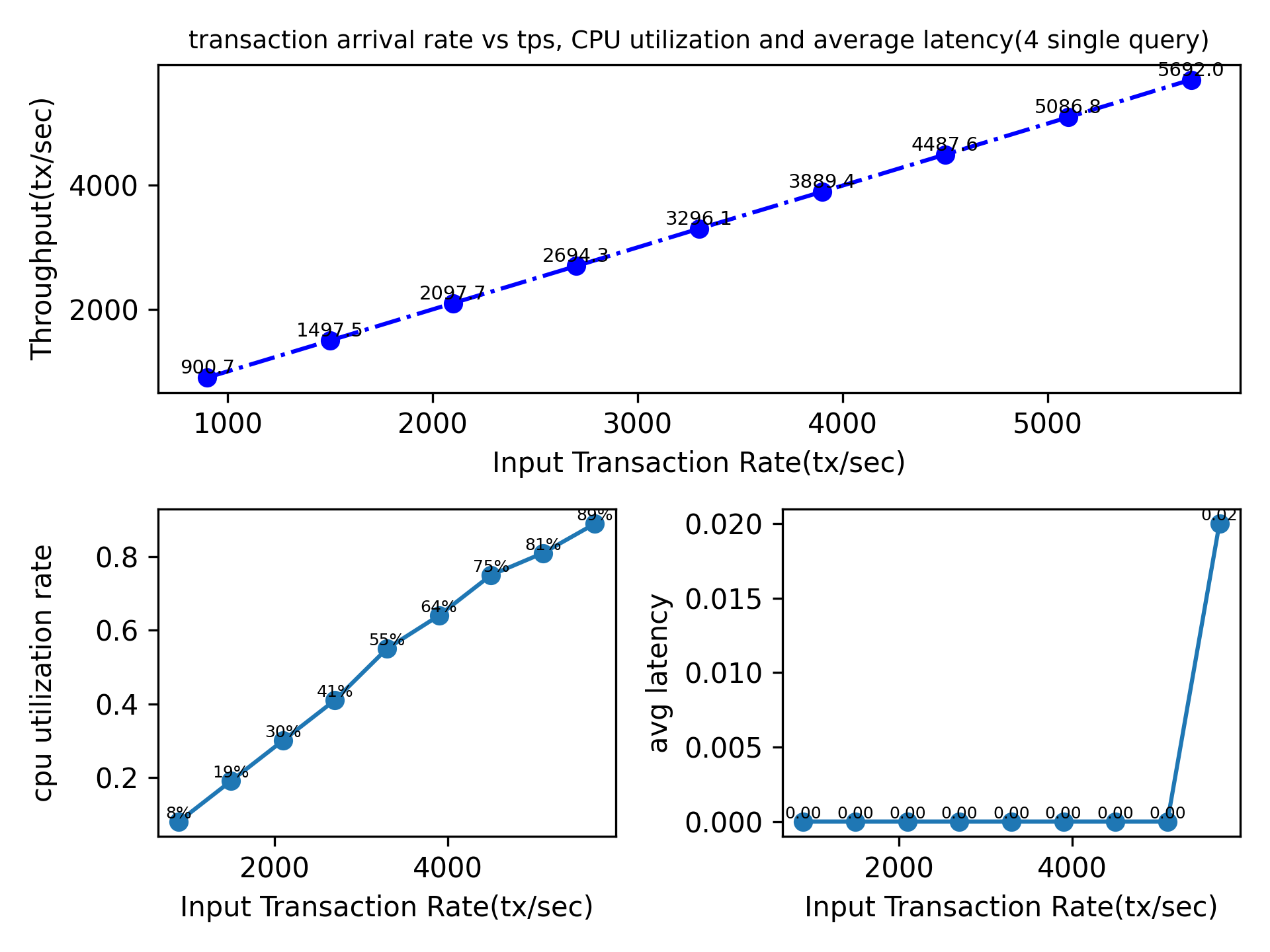}
  \caption{transaction arrival rate vs TPS, CPU utilization and average latency(4 single read)}
  \label{fig:07}
\end{figure}

\begin{figure}[!htbp]
  \centering
  \includegraphics[width=\linewidth]{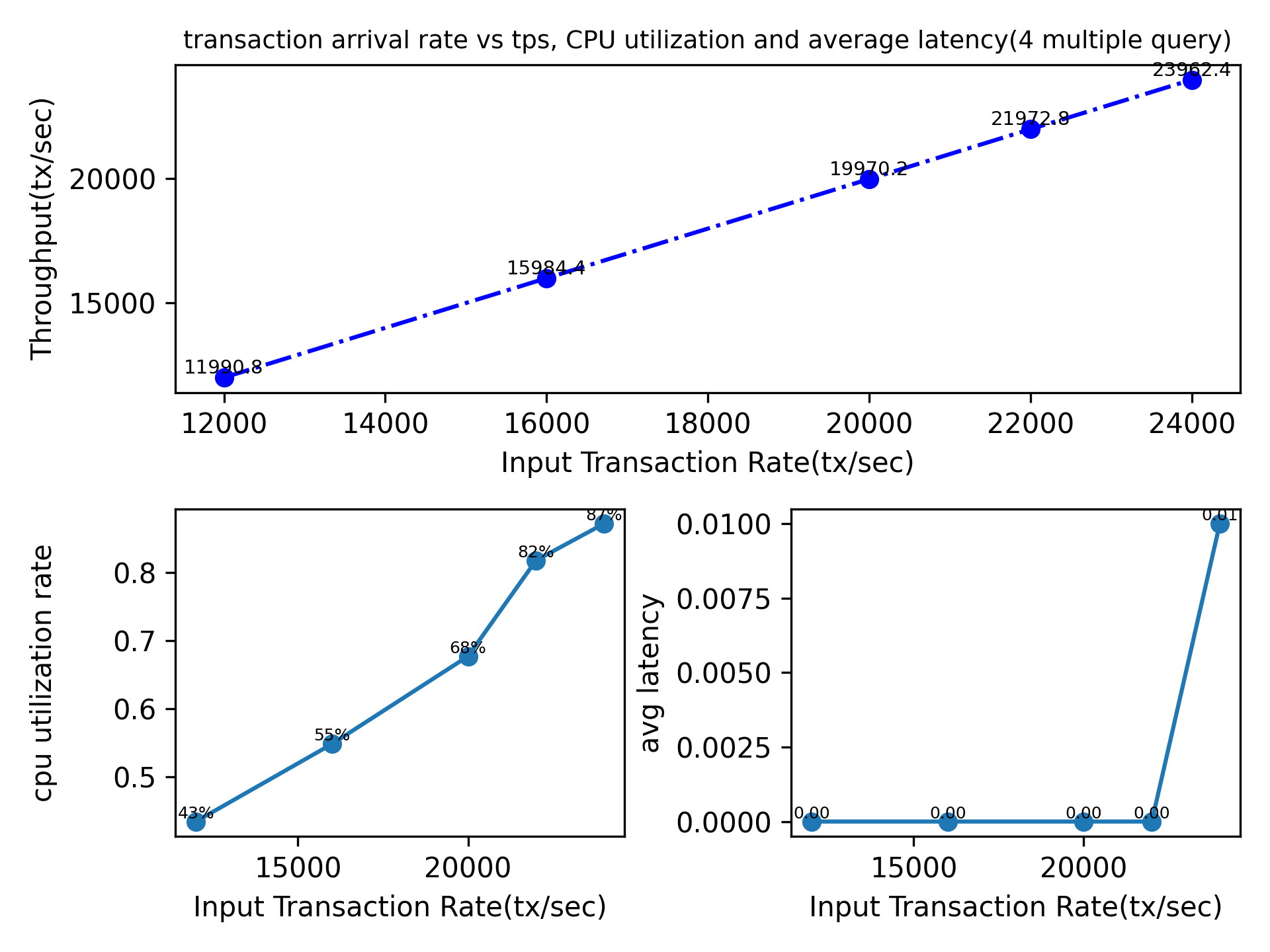}
  \caption{transaction arrival rate vs tps, CPU utilization and average latency(4 multiple read)}
  \label{fig:08}
\end{figure}

We then evaluated the performance of reading transactions on a blockchain consisting of four nodes, as shown in Figure \ref{fig:07}-\ref{fig:08}. we measured the TPS, CPU utilization, and average latency of block queries on single and multiple nodes. we found that the blockchain does not affect the query rate, but the performance bottleneck of the machine does.

\subsubsection{Performance evaluation with Poisson distributed transaction arrival rate model}

We selected four nodes for the evaluation of this experiment. We divided the "read" and "write" into two main groups of experiments. Each large group of experiments is divided into multiple groups of experiments, and the value of $\lambda$ is different for different groups. We measure 5 times in each group of experiments. In each experiment, we send a Poisson distribution as the value of $\lambda$ to inject transactions into the system for ten minutes. And get the corresponding average TPS, latency, and resource usage.

\begin{figure}[!htbp]
  \centering
  \includegraphics[width=\linewidth]{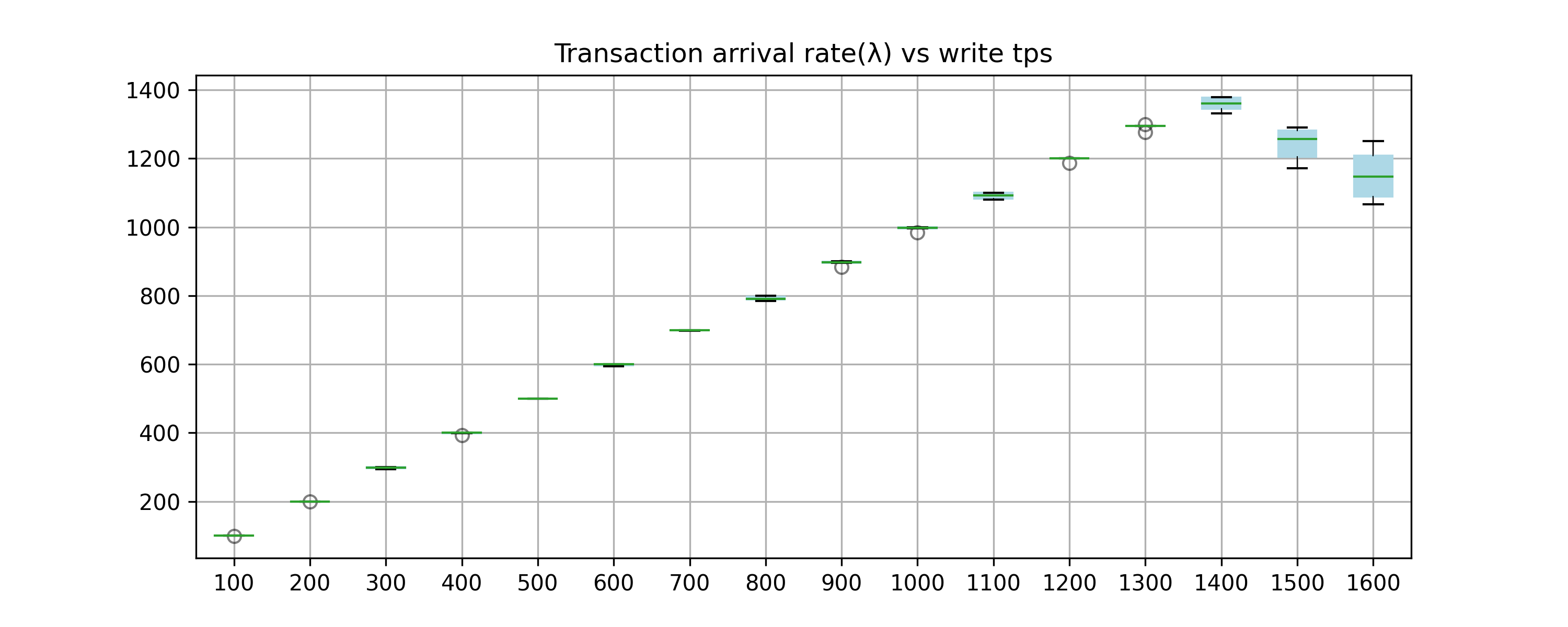}
  \caption{4 node transaction arrival rate(Use the poisson distribution model) vs 4 node write tps}
  \label{fig:09}
\end{figure}

As shown in Figure \ref{fig:09}, the system reaches maximum throughput when $\lambda_\alpha$ = $20,500$ for the query transaction. When the read transactions for four nodes reach $\lambda_\alpha$ $\textgreater$ 20,500, the system performance begins to decline precipitously.

\begin{figure}[!htbp]
  \centering
  \includegraphics[width=\linewidth]{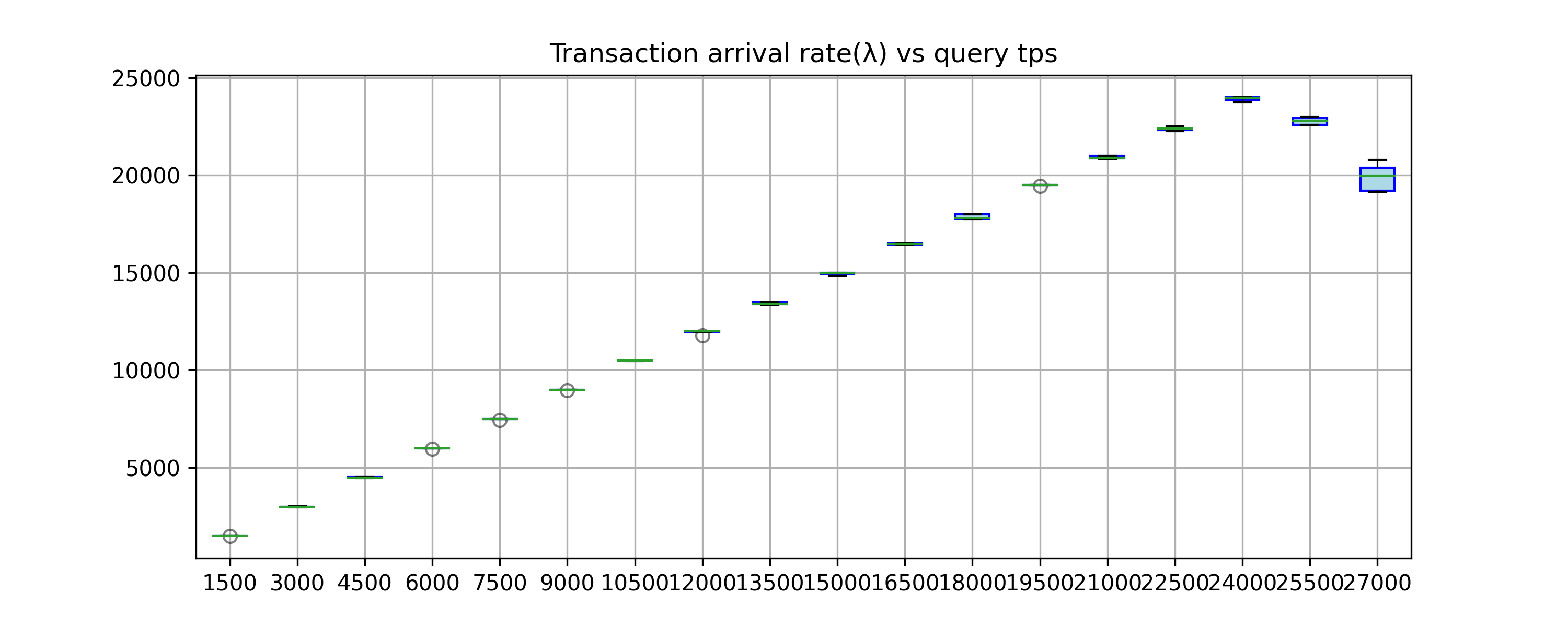}
  \caption{4 node transaction arrival rate(Use the poisson distribution model) vs 4 node multiple read tps}
  \label{fig:10}
\end{figure}

As shown in Figure \ref{fig:10}, the transaction input rate of the system is equal to the system throughput rate (i.e., reaches steady-state). However, when $\lambda_\beta$ exceeds 1400, the system's performance degrades dramatically. Compared to Figure \ref{fig:03}, we observe that the system reaches a steady-state with a value of $\lambda_\beta$ slightly smaller than the peak. There is a chance, according to the Poisson model, that a value greater than $\lambda_\beta$ will occur, causing the system's transaction processing rate to be lower than the transaction arrival rate at some time.

Adopting the methodology provided in Part IV, we analyze whether the performance of the quorum blockchain meets the criteria of the seven major scenarios. Figures \ref{fig:09} and \ref{fig:10} depict the maximum read and write transaction throughputs for the quorum blockchain using the Poisson distribution traffic arrival model with $\lambda_\alpha$ = 20,500 and $\lambda_\beta$ = 1400, respectively. In section IV, we compare the read and write transaction arrival rates to the maximum throughput for the seven scenarios and conclude that our system is well-suited. For example, the write transaction rate for the public key management scenario is $\lambda_\beta$ = 0.0115, and the Quorum blockchain meets all performance requirements for this scenario. The write transaction rate for the AAA case, however, is $\lambda_\beta$ = 8333 and the read transaction rate is $\lambda_\alpha$ = 41665, which surpasses the maximum throughput of the Quorum blockchain. Consequently, this scenario is not suitable for the Quorum blockchain. This problem may be remedied in a number of ways, including by consolidating several transactions into a single one or by scaling the blockchain. We have thoroughly investigated the remaining scenarios and found that they can be accommodated by the Quorum blockchain. The aforementioned study indicates that a properly constructed federated blockchain is able to match the performance requirements of a 6G network scenario.

\section{Conclusion}

6G blockchain, although still in its early stages, has attracted more and more researchers and companies to get involved in investigating it. This article analyzes seven use scenarios of blockchain under 6G and presents a methodology for evaluating the usability of the scenarios using certain blockchain. Concretely, we start by analyzing three why, how, and when in conjunction with seven 6G scenarios and blockchain. After that, we propose a methodology. We may evaluate the usefulness of 6G scenarios with the aid of the methodology. It also does a basic evaluation of the Quorum blockchain's performance. In addition, Quorum is assessed using a traffic model with a Poisson distribution for transaction arrival rates. It calculates the arrival rates for seven scenarios and evaluates the performance of the blockchain system in a real-world environment to assure the blockchain's availability in 6G. Finally, The experimental results show that consortium blockchain with the proper settings may satisfy the performance and scalability requirements of a 6G network.

In conclusion, We attempted to provide a \textbf{Methodology} and \textbf{Complete Experiment} to spur interest and further investigations for subsequent research on blockchain-empowered 6G systems. nevertheless, there are still certain difficulties that are not addressed in our study. 1) Although we have outlined seven 6G possibilities, Other 6G scenarios, such as Telematics and Drones, may also use blockchain. More 6G application scenarios must be explored, and the usability of the scenarios under blockchain must be evaluated; 2)This paper focuses on the usability of blockchain in a single scenario of 6G, but for a real 6G network where all scenarios must be covered, a complete blockchain architecture is required to meet the needs of the entire 6G scenario. the communication and interaction across various situations must also be accomplished inside this framework. Therefore, this effort is crucial for the future.

\bibliographystyle{IEEEtran}
\bibliography{ref}

\vspace{11pt}

\begin{IEEEbiography}[{\includegraphics[width=1in,height=1.25in,clip,keepaspectratio]{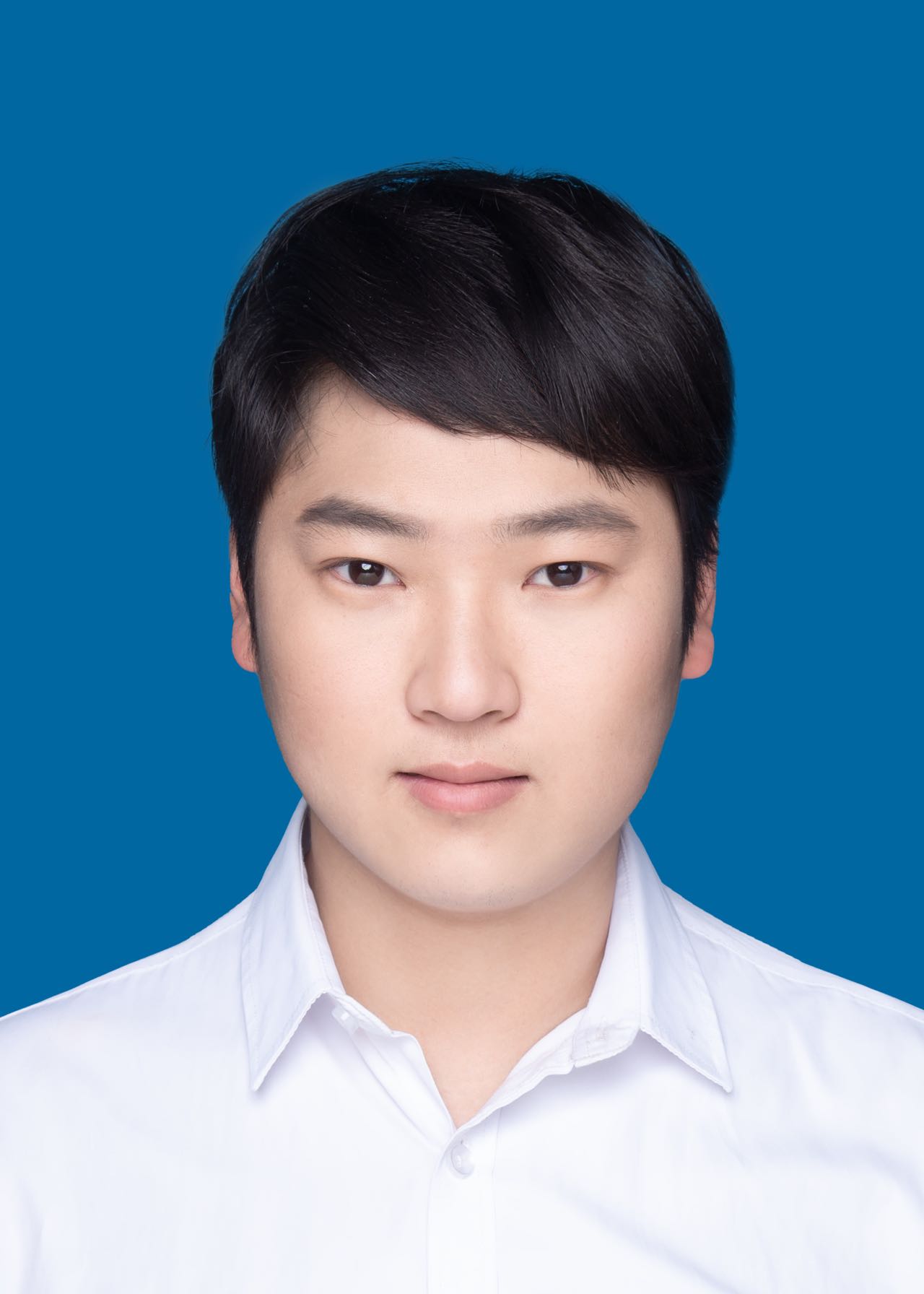}}]{Bo Li}
received the B.S. degree in 2016 from the College of Computer and Information, Anhui Polytechnic University. he is currently pursuing the PhD.degree with the College of Computer Science and Technology, Zhejiang University, Hangzhou, China. His research interests include Blockchain, Cryptology, Cloud Computing and Edge Computing.
\end{IEEEbiography}

\vspace{11pt}

\begin{IEEEbiography}[{\includegraphics[width=1in,height=1.25in,clip,keepaspectratio]{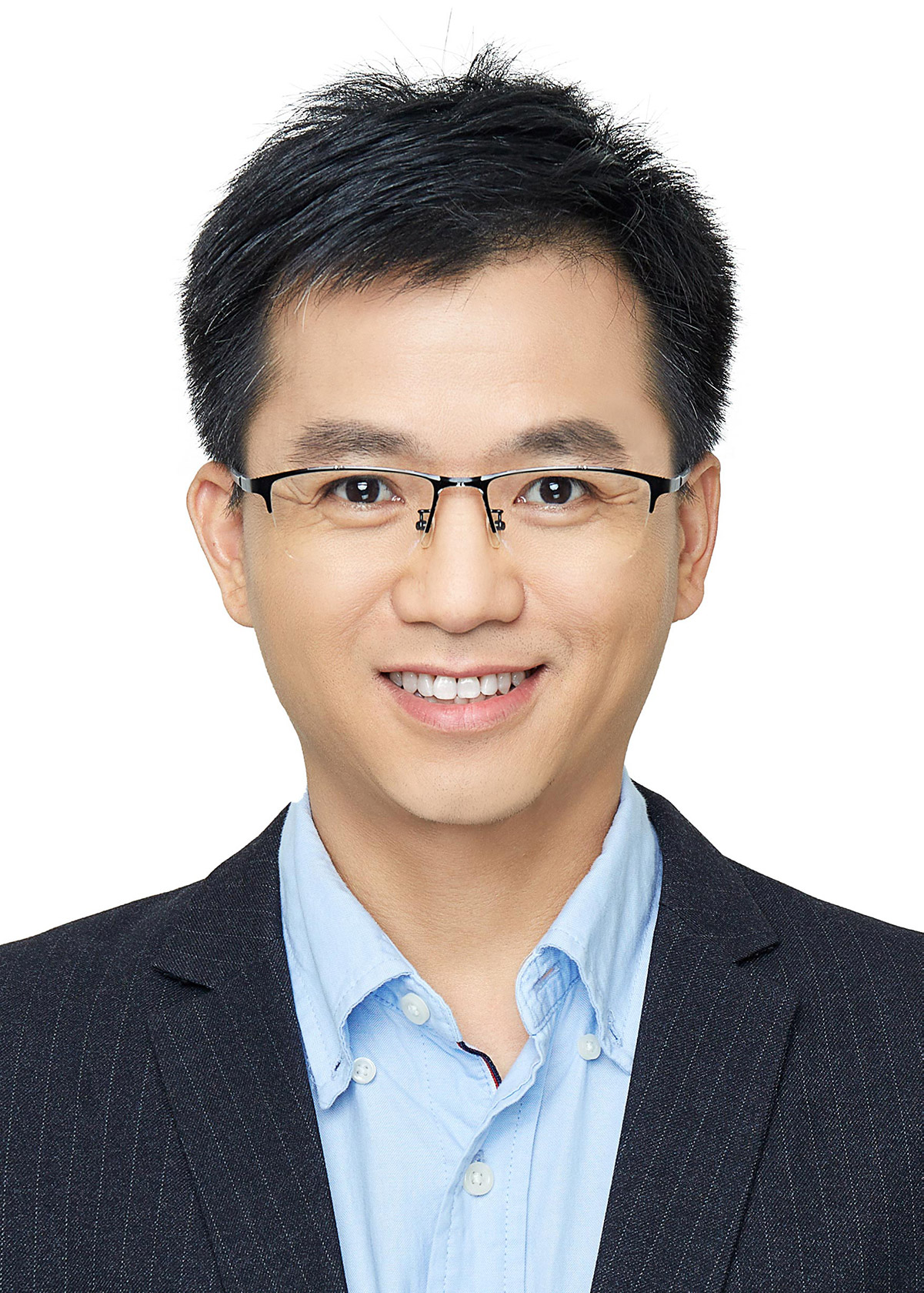}}]{Shuiguang Deng}
is currently a full professor at the College of Computer Science and Technology in Zhejiang University, China, where he received a BS and PhD degree both in Computer Science in 2002 and 2007, respectively. He previously worked at the MIT in 2014 and Stanford University in 2015 as a visiting scholar. His research interests include Edge Computing, Service Computing, Cloud Computing, and Business Process Management. He serves for the journal IEEE Trans. on Services Computing, Knowledge and Information Systems, Computing, and IET CPS as an Associate Editor. Up to now, he has published more than 100 papers in journals and refereed conferences. In 2018, he was granted the Rising Star Award by IEEE TCSVC. He is a fellow of IET and a senior member of IEEE.
\end{IEEEbiography}

\vspace{11pt}

\begin{IEEEbiography}[{\includegraphics[width=1in,height=1.25in,clip,keepaspectratio]{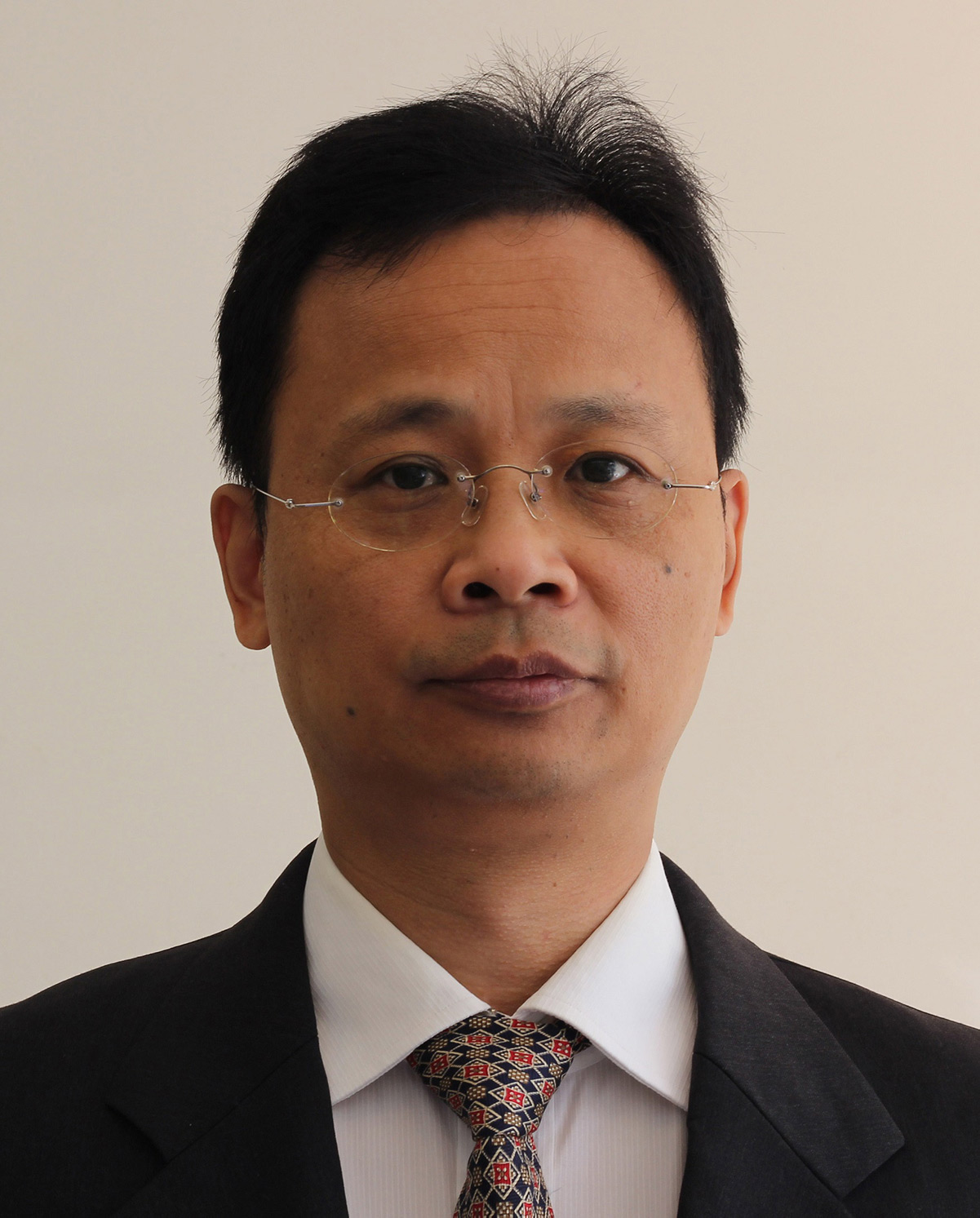}}]{Xueqiang Yan}
is currently a Technology Expert with Wireless Technology Lab, Huawei Technologies. He was a Member of Technical Staff with Bell Labs from 2000 to 2004. From 2004 to 2016, he was the Director of Strategy Department, Alcatel-Lucent Shanghai Bell. His current research interests include future mobile network architecture, edge AI, data analytics, Blockchain and Internet of Things.
\end{IEEEbiography}

\vspace{11pt}

\begin{IEEEbiography}[{\includegraphics[width=1in,height=1.25in,clip,keepaspectratio]{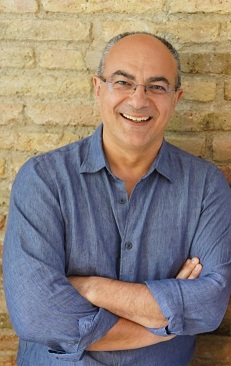}}]{Schahram Dustdar}
is a Full Professor of Computer Science (Informatics) with a focus on Internet Technologies heading the Distributed Systems Group at the TU Wien. He is Chairman of the Informatics Section of the Academia Europaea (since December 9, 2016). He is elevated to IEEE Fellow (since January 2016). From 2004-2010 he was Honorary Professor of Information Systems at the Department of Computing Science at the University of Groningen (RuG), The Netherlands. From December 2016 until January 2017 he was a Visiting Professor at the University of Sevilla, Spain and from January until June 2017 he was a Visiting Professor at UC Berkeley, USA. He is a member of the IEEE Conference Activities Committee (CAC) (since 2016), of the Section Committee of Informatics of the Academia Europaea (since 2015), a member of the Academia Europaea: The Academy of Europe, Informatics Section (since 2013). He is recipient of the ACM Distinguished Scientist award (2009) and the IBM Faculty Award (2012). He is an Associate Editor of IEEE Transactions on Services Computing, ACM Transactions on the Web, and ACM Transactions on Internet Technology and on the editorial board of IEEE Internet Computing. He is the Editorin-Chief of Computing (an SCI-ranked journal of Springer).
\end{IEEEbiography}

\vfill
\end{document}